\documentstyle[prl,aps,twocolumn]{revtex}
\begin{document}
\draft
\preprint{}
\title{Dynamical Nonlinear Optic Coefficients 
from the Symmetrized Density Matrix 
Renormalization Group Method}
\author{ Swapan K. Pati$^{1}$, S. Ramasesha$^{1,2}$, 
Z. Shuai$^{3}$ and J. L. Br{\' e}das$^{3}$} 
\address{\it ${^1}$Solid State and Structural Chemistry Unit,
Indian Institute Of Science, Bangalore 560012 , India.}
\address{${^2}$Jawaharlal Nehru Center for Advanced Scientific
Research,
Jakkur Campus, Bangalore 560064, India.}
\address{${^3}$Centre de Recherche en Electronique et
Photonique Mol{\' e}culaires,
Service de Chimie des Mat{\' e}riaux Nouveaux,
Universit{\' e} de Mons-Hainaut, 
Place du Parc 20, 7000 Mons, Belgium.}
\date{\today}
\maketitle
\begin{abstract}
We extend the symmetrized density matrix renormalization group (SDMRG)
method to compute the dynamic nonlinear optic coefficients for long
chains. By computing correction vectors in the appropriate
symmetry subspace we obtain the dynamic
polarizabilities, $\alpha_{ij}(\omega)$, and third-order polarizabilities  
$\gamma_{ijkl}(\omega,\omega,\omega)$ of the 
Hubbard and "$U-V$" chains in an $all-trans$ polyacetylene geometry, 
with and without dimerization. We rationalize the behavior of $\bar{\alpha}$
and $\bar{\gamma}$ on the basis of the low-lying excitation gaps in the
system.
This is the first study of the dynamics of a fermionic system within the
DMRG framework. 
\end{abstract}

\narrowtext
The dynamic linear and nonlinear response functions of 
finite interacting systems have 
most commonly been obtained from an explicit computation of the
eigenstates of the Hamiltonian and the matrix elements of the appropriate
operators in the basis of these eigenstates\cite{orr}. 
This has been the most
widely used method, particularly in the computation of the dynamic 
Nonliear Optic (NLO)
coefficients of molecular systems and is known as the sum-over-states
(SOS) method. In the case of model 
Hamiltonians, the method that has been widely used to study dynamics
is the Lanczos 
method\cite{recur,dagatto}. The spectral intensity corresponding 
to an operator $\hat O$  is given by

$$ I (\omega) = -{1 \over \pi} Im[<G| \hat O^{\dag}{ 1 \over 
{(E_{0} +z -{\hat H}})} \hat O |G>]  \eqno(1) $$

\noindent where $|G>$ is the ground state eigenvector of the Hamiltonian 
with eigenvalue $E_{0}$, $z=\omega+i\epsilon$, $\omega$ is the frequency at which the
response is sought, $\epsilon$ is the mean life time parameter and
${\hat H}$ is the Hamiltonian. In the Lanczos method, $I(\omega)$ is 
computed as a continued fraction,

$$I(\omega) = -{\frac {1}{\pi}} Im[ \frac {<G|\hat O^{\dag}\hat O|G>} 
 {\displaystyle z-a_{0} - \frac {b^{2}_{1}}{\displaystyle z - a_{1}- 
\frac{b^{2}_{2}}{\displaystyle z - a_{2}- \ldots}}}] \eqno(2) $$

\noindent wherein the coefficients $a_{i}$ and $b_{i}$ 
are respectively the diagonal and the 
off-diagonal matrix elements in the tridiagonal matrix representation of 
the Hamiltonian obtained in the
Lanczos procedure. In this method, there is 
an implicit truncation of the
Hilbert space due to the smaller size of the tridiagonal matrix compared with
the total dimensionality of the space spanned by the Hamiltonian\cite{dagatto}.
Therefore, the dynamic quantities computed by this technique
are approximate even though the ground state obtained is exact. 

In the context of NLO properties\cite{zyss,keiss,baker} of the Hubbard
and extended Hubbard Models it was shown by Soos and Ramasesha that the
model exact dynamical NLO coefficents could be obtained by solving for
correction vectors\cite{soossr}. If we define the 
correction vector $\phi^{(1)}(\omega)$ by the equation
$$ ( \hat H - E_{0} -z ) \phi^{(1)}(\omega) = - {\hat O} 
|G>,  \eqno(3) $$
\noindent then the spectral function, $I(\omega)$, can be expressed as
$$ I (\omega)  =  - \frac {1}{\pi} Im <G| \hat O^{\dag} | \phi^{(1)}(\omega) >  \eqno(4) $$

The correction vector is solved for in the basis of the configuartion 
functions, which is also the basis in which the Hamiltonian matrix is
set-up for obtaining the ground state. Given the ground state
and the correction vector, it is straightforward to compute the spectral
function. This method is quite general and has been employed in the
computations of dynamic NLO coefficients of a wide variety
of Hamiltonians\cite{srzs}. The inhomogeneous linear algebraic 
equations encountered in this method 
involve large sparse matrices and an iterative small matrix algorithm, which
runs parallel to the Davidson algorithm for eigenvalue problems, gives
rapid convergence for the solution of the system of equations\cite{sr}. 

The two correction vectors $\phi^{(1)}_{i}( \omega_{1})$ and
$\phi^{(2)}_{ij}(\omega_{1}, \omega_{2})$ encountered in the computation of
polarizability and third-order polarizability are solved for from the 
following linear equations .
$$( H -E_{0} + \omega_{1}+ i\epsilon)|\phi^{(1)}_{i}(\omega_{1}) = 
{\tilde \mu}_{i}|G> \eqno(5) $$
$$( H -E_{0} + \omega_{2}+ i\epsilon)|\phi^{(2)}_{ij}(\omega_{1}) =
{\tilde \mu}_{i}|\phi^{(1)}_{j}(\omega_{1})> \eqno(6) $$

\noindent where ${\tilde \mu_{i}}$s are the {\it dipole displacement}
matrices and other quantities are as defined in eqn.(1).
The DMRG method\cite{white1,white2} as implemented, readily
provides us with the ground state and the Hamiltonian matrix. 
The matrices of the
dipole operators are constructed in the DMRG scheme by renormalizing
the matrix representations of the dipole operator corresponding to
the left and right parts of the system using
the density matrix eigenvector basis in a way which is completely
analogous to the corresponding Hamiltonian operators for the fragments. 
The matrix representation of the dipole operators for the full system
are obtained as direct products of the fragment matrices analogous to the
way by which the full Hamiltonian matrix is constructed. The 
dipole displacement matrices are obtained by subtracting the 
corresponding components of the ground state dipole moments from the 
diagonal elements of the dipole matrices.

In terms of these correction vectors, the components of the polarizabilities, 
$\alpha_{ij}$, and third-order polarizabilities, $\gamma_{ijkl}$, 
can be written as

$$\alpha_{ij}(\omega)=<\phi^{(1)}_{i}(\omega)|{\tilde \mu}_{j}|G>+
<\phi^{(1)}_{j}(-\omega)|{\tilde \mu}_{i}|G>, \eqno(7) $$ 

$$\gamma_{ijkl}(\omega_{1},\omega_{2}, \omega_{3}) ={\hat P} <\phi^{(1)}_{i}
(\omega_{\sigma})|{\tilde \mu}_{j}
|\phi^{(2)}_{kl}( -\omega_{1} - \omega_{2},-\omega_{1})> \eqno(8) $$ 
\noindent where the operator ${\hat P}$ generates all the permutations: 
$( -\omega_{\sigma},i),
(\omega_{1}, j), ( \omega_{2}, k)$ and $(\omega_{3},l)$ leading to 24
terms for $\gamma_{ijkl}$ with $\omega_{\sigma} = -\omega_{1}
 -\omega_{2} -\omega_{3}$.
The tumbling averaged ${\bar {\alpha}}$ and ${\bar {\gamma}}$ can be defined as
$${\bar {\alpha}} = \frac {1}{3} \sum \limits_{i=1}^{3} \alpha_{ii}~~;
{\bar {\gamma}} = \frac {1}{15} \sum \limits_{i,j =1}^{3} ( 2\gamma_{iijj}+
 \gamma_{ijji}) \eqno(9) $$
\noindent to allow comparison of the calculated NLO response 
with experiments on systems
containing molecules in random orientations\cite{zssr}.

The DMRG method as usually implemented, does not 
exploit all the symmetries
of the system. In the case of model Hamiltonians for polymers, the system
posseses total spin symmetry,  reflection
symmetry about the middle of the chain and in some cases the 
electron-hole symmetry. These symmetries ensure 
that a given correction 
vector spans only a symmetrized subspace and not the entire Hilbert space
of the Hamiltonian of the given system. The correction vector lies in
the symmetry subspace which is connected to the ground state by the electric
dipole operator. For $\omega$ values corresponding to resonance between
the ground state and eigenstates in that particluar symmetry subspace,
$lhs$ of eqns(5) and (6) become singular for $\epsilon=0$ and
present numerical difficulties even when solving them for reasonable
nonzero $\epsilon$ values. However, if we do not exploit the symmetries
of the Hamiltonian, we encounter sigularities in eqns(5) and (6) even
for those $\omega$ values corresponding to eigenstates of the Hamiltonian 
found in other
symmetry subspaces which are not connected to the ground state by
the dipole operator. Therefore, numerically
it would be impossible to obtain the correction vectors at these
frequencies and thereby the associated response of the system. 

For example, in Hubbard chains at intermediate 
correlation strengths, a triplet excited 
state lies below the lowest singlet state in the {\it ionic} 
B subspace\cite{srgal}. The states in 
the {\it ionic} B space are connected to the ground state via one-photon 
transitions. 
The resonances in polarizability are thus expected only at frequencies 
corresponding to the energy levels in the {\it ionic} B space, 
relative to the ground 
state. However, we can not solve for the correction vector using equation (5) 
at an excitation energy corresponding to the the energy of the lowest triplet 
state. Thus the technique of correction vectors will 
not be able to give the complete  
dispersion of the polarizabilities upto the first one-photon resonance,
unless interferences due to spurious intruders such as the triplet states 
are eliminated by suitably block-diagonalizing the Hamiltonian matrix.
This problem of intruders becomes more severe with increasing system size
due to increasing number of intruder states lying below the frequency
corresponding to the first "true" resonance.

We have exploited the electron-hole symmetry, the reflection symmetry
of the polymers and spin parity to block-diagonalise the Hamiltonian\cite{srskp}.
The electron-hole symmetry divides the Hilbert space into {\it ionic} and
{\it covalent} spaces. The ground state is in the covalent space while all
the dipole allowed excitations from the ground state lie in the 
{\it ionic} space. Use of parity conservation divides the Hilbert space
of the Hamiltonian into even and odd parity spaces corresponding to
even and odd total spin states. We have considered a single excitation 
frequency in all our calculations $(\omega=\omega_{1}=\omega_{2} = 
\omega_{3})$ 
and have exploited sparseness of all the matrices to improve upon the
computational efficiency. We have employed the finite size DMRG algorithm
in some cases to check the convergence of the DMRG results. In these cases,
the spatial symmetry is exploited only at the end of the DMRG procedure
when the left and the right density matrices correspond to fragments of the
same size, i.e., at the end of each finite size iteration. 

We have computed the dynamic linear polarizabilities 
({\bf $\alpha (\omega)$}) and 
third-order polarizabilities ({\bf $\gamma (\omega,\omega,\omega)$}) 
corresponding to the third harmonic 
generation (THG), for the Hubbard and "$U-V$" chains of upto 20 sites with 
and without dimerizations, for many values of the parameters of the 
Hamiltonian,
$${\hat H} = \sum \limits_{<ij>,\sigma}[1 - \delta(-1)^{i}]
[ - t({\hat a}^{\dag}_{i,\sigma}
{\hat a}_{j,\sigma} + {\hat a}^{\dag}_{j,\sigma} {\hat a}_{i,\sigma})]
 + U {\hat n}_{i\sigma}{\hat n}_{i-\sigma} +$$ 
$$\sum \limits_{<ij>}[1 - \delta(-1)^{i}][V({\hat n}_{i}-1)
({\hat n}_{j}-1) \eqno(10) $$

\noindent where ${\hat a}^{\dag}_{i,\sigma}  ({\hat a}_{i,\sigma}),
{\hat n}_{i\sigma}$, $t$ and $U$
have their usual meaning.  $V$ is the nearest neighbour 
interaction parameter and the term in $V$ is nonzero only when the
nearest neighbours in question are charged, in which case, the
interaction is repulsive for like charges and attractive for unlike charges.
$\delta$ is the dimerization parameter of the system.
The geometry of the Hubbard and "$U-V$" chains required for the
computation of the NLO coefficients is chosen to correspond to 
$all-trans$ polyacetylene configuration, with a bond angle
of $120^{\circ}$ and bondlength of $1 \AA$, for the uniform
chain. Dimerization ($\delta > 0$) leads to proportionate 
alternation in bond lengths as well as alternation in the
$t$ and $V$ parameters. All
the computations have been carried out at a single frequency corresponding
to an excitation energy of $0.1eV$ and the life-time $\epsilon$ is chosen
to be $0.001eV$.

We have compared the DMRG results with a cut-off of 
$m=200$ (retaining the dominant 200 density matrix 
eigenvectors in the DMRG scheme)
with the {\it model exact}~ {\bf $\alpha$}~ and 
~{\bf $\gamma$}~~ values obtained from the 
correction vector method for chains of upto 12 sites. 
For the 8-site problem, the DMRG calculation is exact 
for this cut-off and the DMRG
results compare with exact results to numerical accuracy. In the case
of the 12 site uniform Hubbard chain with $U/t=4$, the model exact 
${\bar \alpha}$ is $5.343 \times 10^{-24}$ $esu$ 
while the DMRG ${\bar \alpha}$ is $5.293 \times 10^{-24}$ $esu$. 
The model exact ${\bar
\gamma}$ value is $598.3 \times 10^{-36}$ $esu$ while the DMRG 
${\bar \gamma}$ value is $589.3 \times 10^{-36}$ $esu$. The error in the dominant ${\bf \alpha}$ and ${\bf \gamma}$ 
components namely, ${\alpha_{xx}}$ and ${\gamma_{xxxx}}$ is much smaller.
The model exact ${\alpha_{xx}}$ is $14.83 \times 10^{-24}$ $esu$ while the 
DMRG ${\alpha_{xx}}$ is $14.79 \times 10^{-24}$ $esu$. As for 
$\bf {\gamma}$, the model exact ${\gamma_{xxxx}}$  is $2873 \times 10^{-36}$
 $esu$ while the DMRG ${\gamma_{xxxx}}$ is $2870 \times 10^{-36}$ $esu$.

In figs.1, we dispay the dependence of $\bar \alpha$ on the chain length for
different values of $U/t$, for uniform (fig. 1a) and dimerized (fig. 1b) 
Hubbard chains. The polarizability decreases with increasing correlation 
strength in both the dimerized and the uniform chains. For the same chain 
length the uniform chains have, as expected\cite{marder}, higher polarizability than the corresponding 
dimerized chain at every value of $U/t$ we have studied. For the uniform
chain, the average polarizability, $\bar {\alpha}$, for weak correlations 
exhibits a nice power law dependence on chain length with an exponent
of $2.022 \pm 0.001$. However, for stronger correlations, the polarizability 
deviates from a power law dependence and seems to show a size-consistent
variation at longer chain lengths. The chain length at which the
the change over from the power law behavior occurs, systemetically reduces
with increasing $U/t$. As the chain dimerizes, the range over which 
power law behavior is observed decreases (Fig. 1b). Eventually, in
the limit $\delta = 1.0$, we would observe only size-consistent 
dependence as, in this limit  the chain breaks down to noninteracting dimers.

In figs.2, we show the log-log plot of average third-order polarizabilty 
$\bar \gamma$ with the  chain length for chains of upto 20 sites 
with (fig. 2a) zero and (fig. 2b) nonzero bond alternation. The
variation of $\bar {\gamma}$ with chainlength is similar to that
found for $\bar {\alpha}$ both in the dimerized and undimerized cases.
Nonzero $\delta$ leads to a decrease
in the $\bar \gamma$ value and the decrease is smaller for higher 
$U/t$ values. An exponent of $5.570 \pm 0.001$ is found for $\delta=0.0$
and $U/t = 2.0$, while the exponent is $4.00 \pm 0.01$ for $\delta=0.09$
and $U/t = 2.0$. In both these cases, the power-law behavior is observed
upto the maximum chain length we have studied. It is interesting to
note the strong dependence of the exponent on the dimerization parameter.

In the $"U-V"$ model the dependence of the polarizability on the
chain length is very simialr to that in the Hubbard model.
However, the "$U-V$" model is more polarizable, for the chosen model 
parameters and frequency. While the dimerization, $\delta$, decreases the polarizability
of the chain, the nearest neighbour interaction, $V$, increases the
same.  The exponents for $\bar {\alpha}$,
where the power law holds ($U/t =2.0, V/t=1.0$) is $2.320 \pm 0.001$ for the 
uniform chain and $1.64 \pm 0.01$ for the dimerized chain($\delta = 0.09$).

In figs. 3 we present the dependence of $\bar {\gamma}$ on chain length
for various values of $U/t$ for the uniform (fig. 3a) and dimerized 
(fig. 3b) chains, in the $"U-V"$ model with $V/t$ fixed at 1.0. 
There are some very interesting differences in the variation of 
$\bar {\gamma}$ between
the Hubbard and the "$U-V$" chains. The dependence of the $\bar {\gamma}$
of the two chains on $U$ are similar for $U > 2V$. In this 
the $SDW$ regime, the third-order polarizability in
the "$U-V$" model is larger than that in the Hubbard model for 
corresponding chain lengths both with and without dimerization.
However, when $U = 2V$, which is the crossover point from the
$SDW$ to the $CDW$ regime, we find
that the Hubbard chains have higher third-order polarizability than their
counterparts in the "$U-V$" model. This trend is observed
only for $\bar {\gamma}$ and is not seen for the polarizability $\bar
{\alpha}$. To understand this behavior we have studied the 
variation of the $lowest$ $one-photon$ $gap$ (fig. 4a) and the $lowest$ 
$two-photon$ $gap$ (fig. 4b) as a function of the
chain length in the Hubbard model and the $"U-V"$ model at the $SDW/CDW$
transition point. The gap to the lowest one-photon state in the former 
is higher than that in the latter at all 
chain lengths, independent of $\delta$. Therefore the polarizabilities 
of the $"U-V"$ model at the transition point are
higher than that of the Hubbard model. The 
third-order polarizability also has contributions from the two-photon states.
The lowest two-photon state
in the $"U-V"$ model for these parameters is at a higher energy than in 
the Hubbard model for all the chain lengths and $\delta$. 
Consequently, the two-photon contribution to $\gamma$
is higher for the Hubbard model than for the $"U-V"$ model. 

For values of $V/t > U/2t$ ($CDW$ regime), the optical 
gap decreases
sharply and the chosen excitation frequency, $\omega$, of 0.1$eV$ is above the
first three-photon resonance in the system. In this regime, 
$\gamma$ is also found to have a negative sign for sufficiently long 
chains ($N > 10$ sites).

To conclude, we have demonstrated how dynamic NLO responses can be
obtained within the DMRG procedure. We have applied the method to 
Hubbard and "$U-V$" chains in the $CDW$ and $SDW$ regimes.

Acknowledgements: Thanks are due to Prof. H. R. Krishnamurthy for many
useful discussions and Dr. Biswadeb Dutta of JNCASR for system assistance.
The work in Mons is partly supported by the Belgian Government SSTC (P${\hat o}$le
d'Attraction Interunivrsitaire en Chimie Supramol{\' e}culaire et Catalyse), FNRS/FRFC, and IBM Academic Joint Study.

\begin{center}
{\bf Figure Captions}
\end{center}
{\bf Fig.1} \\
Log-Log plot of the average polarizability (${\bar \alpha}$) in $10^{-24}$
$esu$ versus chain length L in $\AA$ for Hubbard chains of upto 20 sites in 
$all-trans$ polyacetylene geometry for (a) $\delta=0$ and (b) $\delta=0.09$,
for three different values of $U/t$. \\
{\bf Fig.2} \\
Plot of the log of average third-order polarizability (${\bar \gamma}$)
in $10^{-36}$ $esu$ versus log of the chain length L in $\AA$, for Hubbard chains,
with three different values of $U/t$ for (a) $\delta=0$ and (b) $\delta=0.09$. \\
{\bf Fig.3} \\
Log-Log plot of the average (${\bar \gamma}$) in $10^{-36}$ $esu$ versus 
chain length L in $\AA$ for "$U-V$" chains for three values of $U/t$ and 
for $V/t=1$ for (a) $\delta=0$ and (b) $\delta=0.09$. \\
{\bf Fig.4} \\
Dependence of (a) one-photon gap (ground state to $1^{1}B^{-}_{u}$) and
(b) two-photon gap ( ground state to $2^{1}A^{+}_{g}$) on $1/N$. $N$ is the
number of sites in the chain.

\end{document}